\documentclass[
10pt,tightenlines,
amsmath,
amssymb,
prl,
twocolumn,
showpacs,
preprintnumbers]{revtex4}

\newcommand{\beq}{\begin{equation}}
\newcommand{\eeq}{\end{equation}}
\newcommand{\bea}{\begin{eqnarray}}
\newcommand{\eea}{\end{eqnarray}}
\newcommand{\bmat}{\begin{pmatrix}}
\newcommand{\emat}{\end{pmatrix}}
\newcommand{\nn}{\nonumber}
\newcommand{\junk}[1]{}
\def\<{\langle}
\def\>{\rangle}
\def\d{\partial}
\def\+{\dagger}

\def\UEM{$U(1)_{EM}~$}

\def\mcoop{m_c}

\begin{document}

\title{Neutron stars as type-I superconductors}
\author{Kirk~B.~W.~Buckley, Max~A.~Metlitski, and Ariel~R.~Zhitnitsky}
\affiliation{Department of Physics and Astronomy,
University of British Columbia,
Vancouver, BC, Canada, V6T 1Z1  }
\date{\today}
\begin{abstract}
In a recent paper by Link, it was pointed out that the
standard picture of the neutron star core composed of a mixture of a
neutron superfluid and a proton type-II superconductor is
inconsistent with observations of a long period precession in
isolated pulsars.   In the following we will show that an
appropriate treatment of the interacting two-component superfluid
(made of neutron and proton Cooper pairs), when the structure of
proton vortices is strongly modified, may dramatically change the
standard picture, resulting in a type-I superconductor. In this case 
the magnetic field is expelled from the superconducting regions of the
neutron star leading to the formation of the intermediate state when
alternating domains of superconducting matter and normal matter 
coexist.
\end{abstract}

\maketitle

The conventional picture of a neutron star is that the extremely
dense interior is mainly composed of neutrons, with a small amount
of protons and electrons in beta equilibrium. The neutrons form
$^3P_2$ Cooper pairs and Bose condense to a superfluid state, while
the protons form $^1S_0 $ Cooper pairs and Bose condense, as well,
to give a superconductor (see e.g. \cite{review} for a review). It
is generally believed that the proton superfluid is a type-II
superconductor, which means that it supports a stable lattice of
magnetic flux tubes in  the presence of a magnetic field. In
addition, the rotation of a neutron star causes a lattice of
quantized vortices to form in the superfluid neutron state,
similar to the observed vortices  that form when superfluid $He$
is rotated fast enough. In a recent paper by Link \cite{link}, it
was pointed out that the precession of the neutron star hints that
this picture may not necessarily be correct. In particular, Link
states that the observed precession of a neutron star does not
allow the proton magnetic flux tubes and neutron vortex lattice to
exist simultaneously, due to the fact that the axis of rotation
and the axis of the magnetic field are not aligned and the fact
that  these two different vortices interact quite strongly.
Furthermore, Link suggests that the conventional picture of a
neutron star as a type-II superconductor may have to be
reconsidered.
One should remark here that the conventional picture of type-II
superconductivity follows from the standard analysis when only a
single proton field is considered. As we shall demonstrate in this
letter, if one takes into account that the Cooper pairs of neutrons
are also present in the system and that they interact strongly with
the proton Cooper pairs,   the superconductor may in fact be type-I
and exhibit the Meissner effect (total expulsion of an external
magnetic field), contrary to the picture that is obtained when
only the proton Cooper pair condensate is accounted for. This
would support the suggestion made by Link \cite{link} that neutron
stars may in fact be type-I superconductors with the
superconducting region not carrying any magnetic flux.

The core of a neutron star is a mixture of neutron and proton
superfluids, as discussed above. In the presence of a magnetic
field, it is well known that the type-II proton superfluid may
form magnetic flux tubes. Inside the core of these vortices, the
proton condensate vanishes, and the core is filled with normal
protons resulting in the restoration of the broken \UEM symmetry.
If the accepted estimates of the proton correlation length and the 
London penetration depth are used, 
then the distant proton vortices repel each other
leading to formation of a stable vortex lattice. This is the
standard picture realized in conventional type-II superconductors.
However, there are many situations where this picture will be
qualitatively modified. For example, if a second field or
component is added, such that there is an approximate $SU(2)$
symmetry between the original and the second fields, it may be
energetically favorable for the second field to condense inside
the vortex core \cite{witten}, resulting in a different
pattern of the vortex-vortex interaction. 
This behavior is known to occur in various systems: 
cosmic strings, high $T_c$ superconductors,
Bose-Einstein condensates, superfluid $^3He$,
and high baryon density quark matter 
(see Refs. \cite{witten,BEC,highTc,krstrings,superk}).

In the case considered in this letter, we have a situation where
there are two condensates, proton and neutron Cooper pairs, both
of which are nonzero in the bulk of the matter.
In what follows 
we shall argue that if the interactions between the proton and neutron
Cooper pairs at small momentum are approximately equal (a precise condition of
``approximately" will be derived below), the vortex-vortex
interaction will be modified and the system will be a type-I
superconductor with the magnetic field completely expelled from the
superconducting regions.
We believe that the approximate symmetry of proton/neutron Cooper 
pair interactions at large distances is somewhat justified
by the original isospin symmetry of bare protons and neutrons,
however this symmetry
is not exactly equivalent to the conventional isotopical $SU(2)$ symmetry.
If we consider a proton vortex (magnetic flux tube) in this case,
the vortex structure is non-trivial, as we will see below. The
core of the proton vortex, where the proton superfluid density
goes to zero, has a neutron superfluid density that is larger than
at spatial infinity, far from the core. Moreover, the size of the
vortex core and the asymptotic behavior of the proton condensate
far from the core are also modified due to the additional neutron
condensate that is present. The most important result of these
effects is that the interaction between distant proton vortices
may be attractive in a physical region of parameter
space leading to type-I behavior: destruction of the
proton vortex lattice and expulsion of the magnetic flux from the
superconducting region of the neutron star. We will now elaborate
on the ideas outlined above.

We start by considering the following effective Landau-Ginsburg
free energy that describes a two component Bose condensed system.
In our system, we have a proton condensate described by the field
$\psi_1$ and a neutron condensate described by the field $\psi_2$.
The $\psi_1$ field with electric charge $q$ (which is twice the
fundamental proton charge, $q=2|e|$) interacts with the gauge field ${\bf
A}$, with ${\bf B} =\nabla \times {\bf A}$. The two dimensional
free energy reads (we neglect the dependence on third direction
along the vortex):
\bea 
\label{dimfree} 
{\cal F} &=& \int d^2 x
    [\frac{\hbar^2}{2 \mcoop}(|(\nabla -   \frac{i q}{\hbar c}{\bf A})
    \psi_1|^2
       + |\nabla \psi_2|^2) \nn\\&+& \frac{{\bf B}^2}{8 \pi} 
	+ V(|\psi_1|^2,|\psi_2|^2) ],
\eea
where $\mcoop = 2 m$ and $m$ is the mass of the nucleon.
Here we have moved the effective mass difference of the proton and
neutron Cooper pairs onto the interaction potential $V$. In the
free energy given above, we have ignored the term coupling the proton
and neutron superfluid velocities, which gives rise to the
Andreev-Bashkin effect \cite{Bashkin}, as it is not important in
our  discussions. 
Indeed, the relevant term in the free energy can be represented 
as $\sim \int d^3 x \vec{v}_1 \cdot \vec{v}_2$,
where $ \vec{v}_1$ and $\vec{v}_2$
are   velocities of the superfluid components.
For neutron stars which do not rotate (the case
which is considered in this paper)
$\vec{v}_2=0$, and the effect obviously vanishes. We expect that due to 
the 
small density of the neutron vortices (compared to the density
of the proton vortices) the effect is still negligible 
for most of the flux tubes in a rotating star as well. 
The effect could be important
only for a few of the flux tubes situated close to a neutron vortex 
core, where $\vec{v}_2$ strongly deviates from the constant value at  
interflux distance scales.
 
We have also ignored the fact that the neutron
condensate has a non-trivial $^3P_2$ order parameter as only the
magnitude of the neutron condensate is relevant to the effect
described below. 
The free energy (\ref{dimfree}) only describes
large distances and it does not describe the gap structure on the Fermi
surfaces, only the superfluid component of the protons and neutrons.

The free energy (\ref{dimfree}) is invariant under a $U(1)_1 \times
U(1)_2$ symmetry associated with respective phase rotations of
fields $\psi_1$ and $\psi_2$, which corresponds to the
conservation of the number of Cooper pairs for each species of
particles. Moreover, we know that the free energy (\ref{dimfree})
describes particles interacting via the strong nuclear force and,
therefore, must be approximately invariant with respect to the
$SU(2)$ isospin symmetry. 
Therefore, we expect that while the Fermi surfaces for protons and neutrons
are very different and the gap equations are very different, the
interaction between different Cooper pairs at small momentum must not be very different 
(the asymmetry must be proportional to $(m_d - m_u)$). This asymmetry is expressed in terms
of different scattering lengths of Cooper pairs for each species.
Thus, we assume, the interaction
potential $V$ can be approximately written as
$V(|\psi_1|^2,|\psi_2|^2) \approx U(|\psi_1|^2 + |\psi_2|^2)$. In
reality this  symmetry is explicitly slightly broken, and the
potential $V$ has a minimum at $|\psi_1|^2 = n_1, |\psi_2|^2 =
n_2$ where  $n_1$ and $n_2$  are the proton and neutron Cooper pair densities.
 Hence in the ground state,
$|<\psi_i>|^2 = n_i, \, i=1,2$, and both $U(1)$ symmetries are
spontaneously broken. An important quantity for the analysis that
follows will be the ratio of proton to neutron Cooper pair density,
$\gamma \equiv n_1 / n_2$. A typical value of $\gamma$ in the core
of a neutron star is $5-15\%$, thus, in our numerical estimates below
 we will often use the limit $\gamma \ll 1$, though our qualitative results
do not depend on this parameter 
\footnote{One should remark
here that the strong deviation of $\gamma$ from $1$ does not imply
a large difference in the interaction and Bose chemical potentials
between different species of
particles. This is in contrast with Fermi systems where large difference 
in densities does imply a large difference in Fermi chemical potentials.}.

Now let's investigate the structure of proton vortices, which
exist due to the spontaneous breaking of the $U(1)_1$ symmetry.
Such vortices are characterized by the phase of the $\psi_1$ field
varying by an integer multiple of $2 \pi$ as one traverses a
contour around the core of the vortex. 
By continuity, the field
$\psi_1$ must vanish in the center of the vortex core. Up to this point,
it has been assumed that the neutron order parameter $\psi_2$
will remain at its  expectation value in the vicinity of the
proton vortex. As we have already remarked, this is not the case in many
similar systems, see e.g.  
\cite{witten,BEC,highTc,krstrings,superk}).

So, anticipating a non-trivial behavior of the neutron field
$\psi_2$, let's adopt the following cylindrically symmetric ansatz
for the fields describing a proton vortex with a unit winding
number:
\beq \label{Ansatz}
\psi_1 = \sqrt{n_1}f(r)e^{i \theta},
\, \psi_2 = \sqrt{n_2}g(r), \, {\bf A} = \frac {\hbar c}{q}
\frac{a(r)}{r} \hat{\theta}
\eeq
where $(r,\theta)$ are the
standard polar coordinates. Here we assume that the proton vortex
is sufficiently far from any rotational neutron vortices, so that
any variation of $\psi_2$ is solely due to the proton vortex. The
functions $f$, $g$, and $a$ obey the following boundary
conditions: $f(0) = 0$, $f(\infty) =1$, $g'(0) = 0$, $g(\infty) =
1$, $a(0) = 0$, and $a(\infty) =1$. We see that the fields
$\psi_1$ and $\psi_2$ approach their  expectation values at
$r = \infty$.

We wish to find the asymptotic behavior of fields $\psi_1$,
$\psi_2$, and ${\bf A}$ far from the proton vortex core, as this
will determine whether distant vortices repel or attract each
other. The asymptotic behavior can be found analytically by
expanding the fields defined in (\ref{Ansatz}):
\beq
f(r) = 1 + F(r), \, g(r) = 1 + G(r), \, a(r) = 1 - r S(r)
\eeq
so that far
away from the vortex core, $F, G, r S \ll 1$ and $F, G, S
\rightarrow 0$ as $r \rightarrow \infty$. This allows us to
linearize far from the vortex core the equations of motion
corresponding to the free energy (\ref{dimfree}) to obtain:
\bea
\label{FG} (\frac{\d^2}{\d r^2} + \frac{1}{r}\frac{\d}{\d r}) \bmat F \\
G \emat = {\bf M} \bmat F \\ G \emat \\ \label{S} S'' +
\frac{1}{r} S' - \frac{1}{r^2} S = \frac{1}{\lambda^2} S
\eea
where the London penetration depth $\lambda = \sqrt{\mcoop c^2/4 \pi
q^2 n_1}$. Here all derivatives are with respect to $r$, and the
matrix ${\bf M}$ mixing the fields $F$ and $G$ is,
\beq
{\bf M} = \frac{4 \mcoop}{\hbar^2} \bmat V_{11} & V_{12} \\
V_{21} & V_{22} \emat \bmat n_1 & 0 \\
0 & n_2 \emat
\eeq
where the derivatives 
$V_{ij}\equiv \partial^2 V/(\partial|\psi_i|^2
\partial|\psi_j|^2)$ are evaluated at $|\psi_i|^2 = n_i$. 
Here we assume that $S^2 \ll F, G$,
i.e. the superconductor is not in the strong type-II regime (this
is justified since we are only attempting to find the boundary
between type-I and type-II superconductivity). The solution to Eq.
(\ref{S}) is known to be:
$S = \frac{C_A}{\lambda} K_1(r/\lambda)$
where $K_1$ is the
modified Bessel function and $C_A$ is an arbitrary constant. The
remaining equation (\ref{FG}) can be solved by diagonalizing the
mixing matrix ${\bf M}$. In previous works the influence of the
neutron condensate on the proton vortex was neglected, which
formally amounts to setting the off-diagonal term $V_{12}$ in
${\bf M}$ to $0$. In that case, one can assume that the neutron
field remains at its  expectation value, i.e. $G = 0$, to
obtain,
$F = C_F K_0(\sqrt{2} r/\xi) $
where $\xi = \sqrt{\hbar^2/2 \mcoop n_1
V_{11}}$ is the correlation length of the proton superconductor
and $K_0$ is the modified Bessel function.
It is estimated that $\lambda \sim 80$~fm and $\xi \sim 30$~fm \cite{link},
which leads to $\kappa = \lambda / \xi \sim 3$ for 
the Landau-Ginzburg parameter. 
As is known from conventional
superconductors, if $\kappa > 1/\sqrt{2}$, distant vortices repel
each other leading to type-II behavior. This is the standard
picture of the proton superconductor in neutron stars that is
widely accepted in the astrophysics community.

However, the standard procedure described above is inherently
flawed since the system exhibits an approximate $U(2)$ symmetry,
which forces approximate equality of second partial derivatives,
$V_{11} \approx V_{22} \approx V_{12}$. This makes the mixing
matrix ${\bf M}$ nearly degenerate. The general solution to Eq.
(\ref{FG}) is:
\beq \label{soln} 
\bmat F \\ G \emat = \sum_{i=1,2}
C_i K_0(\sqrt{\nu_i} r) {\bf v_i}
\eeq
where $\nu_i$ and ${\bf
v_i}$ are the eigenvalues and eigenvectors of matrix ${\bf M}$,
and $ C_i$ are constants to be calculated by matching to the
solution of the original non-linear equations of motion. In the
limit $\gamma = n_1/n_2 \ll 1$ and $\epsilon= (V_{11} V_{22} -
V_{12}^2)/V_{ij}^2 \ll 1$
 one can estimate :
 \beq \nu_1 \simeq
\frac{2 \epsilon}{\xi^2},~~ \nu_2 \simeq \frac{2}{\gamma \xi^2},~~
 {\bf
v_1} \simeq (-1, \gamma),~~ {\bf v_2} \simeq (1,1).
\eeq
The physical meaning of solution (\ref{soln}) is simple: there are
two modes in our two component system. The first mode describes
fluctuations of relative density (concentration) of two components
and the second mode describes fluctuations of overall density of
two components. Notice that $\nu_1 \ll \nu_2$, and hence the
overall density mode has a much smaller correlation length than
the concentration mode. Therefore, far from the vortex core, the
contribution of the overall density mode can be neglected, and one
can write:
 \beq
 \label{assymFG}
\bmat F \\ G \emat (r\rightarrow\infty)
 \simeq C_1 K_0(\sqrt{2 \epsilon} r/\xi) \cdot \bmat -1 \\
 \gamma \emat
\eeq
The most important result of the above discussion is that the
distance scale over which the proton and neutron condensates tend
to their  expectation values near a proton vortex is of
order $\xi/\sqrt{\epsilon}$ - the correlation length of the
concentration mode. Since $\epsilon \ll 1$, this distance scale
can be much larger than the proton correlation length $\xi$, which
is typically assumed to be the radius of the proton vortex core.

We have also verified \cite{BMZ} the above results numerically by
solving the equations of motions corresponding to (\ref{dimfree})
with a particular choice of the approximately $U(2)$ symmetric
interaction potential $V$.
Our numerical
results support the analytical calculations given above. Namely,
we find that the magnitude of the neutron condensate is slightly
increased in the vortex core, the radius of the magnetic flux tube
is of order $\lambda$ and the radius of the proton vortex core is
of order $~\xi/\sqrt{\epsilon}$.

Now that we know the approximate solution for the proton vortex,
we will proceed to look at the interaction between two proton
vortices that are widely separated. If the interaction between two
vortices is repulsive, it is energetically favorable for the
superconductor to organize an Abrikosov vortex lattice with each
vortex carrying a single magnetic flux quantum. As the magnetic
field is increased, more vortices will appear in the material.
This is classic type-II behavior. If the interaction between two
vortices is attractive, it is energetically favorable for $n$
vortices to coalesce and form a vortex of winding number $n$,
which is expelled from the sample. This is type-I behavior.
Typically, the Landau-Ginzburg parameter $\kappa = \lambda/ \xi$
is introduced. In a conventional superconductor, if $\kappa <
1/\sqrt{2}$ then the superconductor is type-I and vortices
attract. If $\kappa
> 1/ \sqrt{2}$ then vortices repel each other and the
superconductor is type-II. As mentioned above, the typical value
for a neutron star is $\kappa \sim 3$, so we would naively expect
that the proton superfluid is a type-II superconductor.


 Now we
will present three different calculations supporting our claim
that for the typical parameters of a neutron star
the proton superconductor may be type-I rather than type-II. 
First of all, we follow the method suggested originally in
\cite{speight} to calculate the force between two widely separated
vortices. The idea of this method is to model distant vortices as
point sources in a free theory, which accurately describes the
behavior of fields far from the vortex cores. The methods of
\cite{speight} were subsequently applied in \cite{mack} to the
case similar to ours, the interaction of two widely separated
vortices that have nontrivial core structure.
Using the asymptotic field solutions 
found above, we follow the procedure of \cite{mack} to obtain the
following expression for the interaction energy per unit vortex
length of two distant parallel vortices:
\beq 
\label{potential} 
U(d) \simeq \frac{2 \pi \hbar^2
n_1}{\mcoop}(C_A^2 K_0(d/\lambda) - C_1^2 K_0(\sqrt{2 \epsilon}d/\xi))
\eeq
where $d \rightarrow\infty$
is the separation between the two vortices. We see that
if the first term in $U$ dominates as $d \rightarrow \infty$ then
the potential is repulsive, otherwise, if the second term
dominates the potential is attractive. 
We introduce the new dimensionless parameter 
$\kappa_{np} = \sqrt{\epsilon} \lambda/\xi$ 
into our description. In terms of this parameter, if
$\kappa_{np} 
< 1/\sqrt{2}$, then vortices attract
each other and the superconductor is type-I; otherwise, vortices
repel each other and the superconductor is type-II. 
Therefore, our parameter $\kappa_{np} =
\lambda/\delta = \sqrt{\epsilon} \lambda/\xi$ should be considered
as an effective Landau-Ginzburg parameter, which determines the
boundary between the type-I and type-II proton superconductivity.
Due to the importance and far reaching consequences of this
result, we have also calculated the vortex-vortex  interaction
energy in a more direct way following \cite{Kramer}; this
calculation \cite{BMZ} produced the same result (\ref{potential})
as the above procedure, therefore confirming our picture.
Our third check of the main result that for relatively
small $\epsilon$ the superconductor in the neutron stars may be,
in fact, type-I is based on the macroscopical calculation of the
critical magnetic fields.
Usually one calculates the critical  magnetic fields $H_c$ and
$H_{c2}$. 
We have calculated \cite{BMZ} the
critical magnetic fields $H_c$ and $H_{c2}$ corresponding to the
free energy (\ref{dimfree}) with a particular choice of $V$ as a
quadratic polynomial in $|\psi_i|^2$. The boundary between type-I
and type-II superconductivity obtained using this procedure
matches the results of our inter-vortex force calculation
presented above.

The most important consequence of this letter is that whether
proton superconductor is type-I or II depends strongly on the
magnitude of the $SU(2)$ asymmetry parameter $\epsilon$.
Specifically, we find that the superconductor is type-I when
$\kappa_{np} = \sqrt{\epsilon} \lambda/\xi < 1/\sqrt{2}$, and
type-II otherwise. This result is quite generic, and not very
sensitive to the
 specific details  of the interaction potential $V$.
In particular, when $\epsilon \rightarrow 0$ the superconductor is
type-I. The parameter $\epsilon$ is not known precisely; the
corresponding  microscopical calculation would require the
analysis of the scattering lengths (amplitudes at small momentum)
of Cooper pairs for different
species. We can roughly estimate this parameter  as being
related to the original $SU(2)$ isospin symmetry breaking
$\epsilon \sim (m_d - m_u)/\Lambda_{QCD} \sim 10^{-2}$. If this is
assumed to be the value of $\epsilon$, we estimate $\kappa_{np}
=\sqrt{\epsilon} \lambda/\xi \sim 0.3 < 1/\sqrt{2}$, which
corresponds to a type-I superconductor. From these crude
estimates, we see that it is very likely that neutron stars are
type-I superconductors with the superconducting region devoid of
any magnetic flux, as was originally suggested in \cite{link} to
resolve the inconsistency with observations of long period
precession in isolated pulsars. 
If this is the case, some of the
explanations of glitches \cite{glitchflux} 
(sudden changes of the neutron star's rotational frequency, see 
\cite{glitchflux} for the original explanation of glitches), 
which assumes a type-II proton superconductor, have to be reconsidered.  
Type-I superconductivity does not imply total expulsion of the magnetic
field: the core structure could be composed of alternating domains
of superconducting matter and normal matter \cite{BMZ}.

We are grateful to M. Prakash, M. Alford, and J. Lattimer for
bringing reference \cite{link} to our attention. This work was
supported in part by the Natural Sciences and Engineering Research
Council of Canada.

\end{document}